\documentstyle[prb,aps,epsfig]{revtex}
\begin{document}
\input{psfig} 
\title{Recurrent oligomers in proteins --
an optimal scheme reconciling accurate and concise backbone
representations in automated folding and design studies}

\author{Cristian Micheletti$^1$, Flavio Seno$^2$ and Amos Maritan$^1$}
\address{(1) International School for Advanced Studies and INFM,
Via Beirut 2, 34014 Trieste, Italy\\
and the Abdus Salam International Centre for Theoretical Physics}
\address{(2) INFM-Biophysics, Dipartimento ``G. Galilei'' , Via
Marzolo 8, 35100 Padova, Italy}
\date{\today}
\maketitle
\begin{tighten}
\begin{abstract}
A novel scheme is introduced to capture the spatial correlations of
consecutive amino acids in naturally occurring proteins. This
knowledge-based strategy is able to carry out optimally automated
subdivisions of protein fragments into classes of similarity. The
goal is to provide the minimal set of protein oligomers (termed
``oligons'' for brevity) that is able to represent any other
fragment. At variance with previous studies where recurrent local
motifs were classified, our concern is to provide simplified protein
representations that have been optimised for use in automated folding
and/or design attempts. In such contexts it is paramount to limit the
number of degrees of freedom per amino acid without incurring in loss
of accuracy of structural representations. The suggested method finds,
by construction, the optimal compromise between these needs. Several
possible oligon lengths are considered. It is shown that meaningful
classifications cannot be done for lengths greater than 6 or smaller
than 4. Different contexts are considered were oligons of length 5 or
6 are recommendable. With only a few dozen of oligons of such length,
virtually any protein can be reproduced within typical experimental
uncertainties. Structural data for the oligons is made publicly
available.
\end{abstract}
\end{tighten}

\tighten

\section{Introduction}

One of the most fundamental and still unsolved problems in biology is
the elucidation of the folding process, that is how a protein sequence
undergoes the structural rearrangements that eventually lead to the
biologically active conformation (believed to be the free energy
minimum) \cite{Anfinsen73}. Since the early studies of Levinthal it
was clear that the dynamics of folding to the native state could not
be governed by mere random processes \cite{Lev}; indeed modern folding
theories explain fast folding processes by invoking
nucleation-condensation mechanisms or funnel-like energy landscapes
\cite{funnel,funnel2} that dramatically reduce the space of visited
conformations \cite{Karplus1,Karplus2,Pit,Micheletti99b}. Topologic,
steric and chemical features are so effective in reducing the space of
viable conformations that even at a local level, only few degrees of
freedom per amino acid are observed. This fact, originally observed by
Ramachandran \cite{Ram} has been lately used in a variety of numerical
schemes. In these approaches proteins are modeled as chains of one or
two interacting centers (representing individual amino acids) with a
limited set of local degrees of freedom, such as torsion angles or
Cartesian positions, chosen to provide optimal compromises between
accurate representation and number of degrees of freedom
\cite{CJ,Levitt,PL96}. These models appear excellent from many points of
view with the exception that they fail to capture correlations between
torsion angles along the peptide chain.

In this paper we address this problem and propose an optimal way to
extend the original idea of Ramachandran of limiting the degrees of
freedom of individual residues to strings of consecutive amino acids,
showing that they are far from independent. Indeed, their correlations
are so strong that, as originally pointed out in a paper by Alwyn
Jones and Thirup \cite{JT86}, it is possible to construct a small data
bank of protein fragments that can be used as elementary building
blocks to reconstruct virtually all native protein structures. We
start by following the seminal idea of Unger et al. \cite{Unger} that
oligomers of a given length found in a coarse grained representation
(such as $C_{\alpha}$ coordinates) of native structures do not vary
continuously but they gather in few clusters. Each of these can be
represented by a single element (that we term ``oligon'') that
optimally catches the geometrical and topological properties of the
entire basin.

Our approach differs from previous work on the classification of
structural fragments \cite{Unger,roomana,roomanb} in that the
procedure we follow to select the oligons has been explicitly
optimised for use in fully automated contexts, especially folding and
design attempts
\cite{Micheletti99b,Pabo83,Quinn94,a4,a6,finger,Micheletti98a,Micheletti98b,annals,Seno96,mayo}.
Indeed, such attempts are commonly framed within numerical problems of
minimizing suitable functionals (such as energy scoring functions) in
structure space. The addition of local constraints reduces drastically
the space of viable structures and is undoubtedly a desired feature
allowing to keep to a minimum the side-effects of using imperfect
parametrizations of the free energy or imperfectly known interaction
potentials
\cite{MJ85,MJ96,sippl1,sippl2,Crippen91,a3,Seno98,Seno98b,du,Settanni,bast_vend}.

The selection strategy we propose is free of subjective inputs or
biases and exploits the full knowledge-based information intrinsic in
our data-bank of non-redundant protein structures. An appealing
feature of the suggested method is that representative fragments are
singled out in order of importance, that is according to the frequency
in which they appear in natural proteins. We carry out a series of
thorough checks and validations of the clustering strategy and show
that the optimal sets of oligons do not suffer from finite-size
effects of the data bank. It is shown that the optimal representatives
have length equal to 5 or 6 and that with only a few tens of them it
is possible to fit virtually any protein within about 1.0 \AA\
co-ordinate root mean square deviations (cRMS) per amino acid.  Some
of the ramifications of this study are discussed and outlined through
preliminary investigations in Section \ref{sec:disc}. The optimal sets
of oligons presented and discussed here are made publicly available at
http://www.sissa.it/\~{}michelet/prot/repset.

\section{Methods and Results}

The first step in the creation of a set of optimal representatives is
the set up of a sufficiently large data bank of protein structures.
Such data bank should cover as best as possible the variety of
distinct protein structures observed in nature. At the same time it is
important to eliminate correlations and biases in the data bank
resulting, for example, from structural homology \cite{lesk}
For
these reasons we compiled our data bank by choosing 75 single-chain
proteins from a carefully compiled list of non-redundant structures
\cite{Settanni}.

The proteins, listed in Table \ref{tab:learn} were chosen from the
SCOP database of non-redundant single-chain proteins covering the most
common families: all $\alpha$, all $\beta$, $\alpha\beta$ and $\alpha
+ \beta$  and the most common chain lengths. This
ensures that, {\em a priori}, the selected structures represent a
broad spectrum of structural instances with the least bias or
redundancy. As discussed later, the results confirm {\em a
posteriori}, that the size and quality of the data-bank was sufficient
for all practical purposes. Each of the proteins of Table
\ref{tab:learn} was partitioned in all the possible fragments of $l$
consecutive residues. We considered values of $l$ ranging from 3 to
10, for which there are 10936 to 10411 distinct fragments. As in
previous studies involved with structural classifications, we retained
only the $C_\alpha$ coordinates of each fragment
\cite{Unger,roomana,roomanb,a1}.

This approach, is practical and consistent with the idea of having an
optimal, but schematic representation of structures. Moreover it is
``reversible'' to a great extent since the whole peptide atomic
geometry can be recovered from the mere knowledge of $C_{\alpha}$
co-ordinates \cite{back}. In turn, if needed, optimal side-chain
rotamer positions could be satisfactorily obtained by exhaustive or
stochastic methods \cite{finger,mayo}.

\subsection{Theory: clustering algorithm}

\begin{figure}
\centerline{\psfig{figure=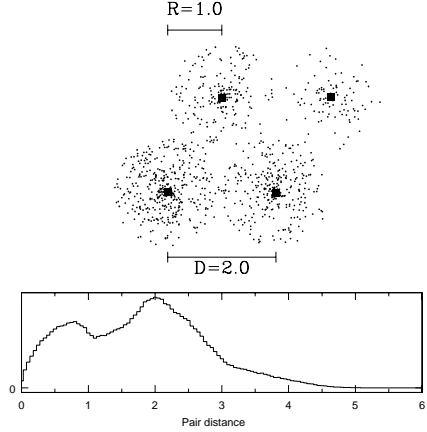,width=2.5in}}
\caption{Illustration of the cluster procedure. 1000 points have been
randomly assigned to cluster of different size but equal radius $R=1$
(arbitrary units). The centers of contacting clusters are at distance
$D=2$. The filled squares correspond to the location of the cluster
centers identified by our procedure. The inset shows the histogram of
distances between any pair of points.}
\label{fig:cluster}
\end{figure}

The goal pursued in this work is to provide a synthetic, but
exhaustive, classification of inequivalent local structural motifs to
be used in contexts where a broad exploration of the space of viable
protein structures is concerned. Hence, the approach pursued here
differs from studies aimed at selecting a restricted number of motif
classes to be used in homology modelling or automated
recognition/classification of secondary motifs
\cite{Unger,roomana,roomanb,a1,a2,a5,a7,a9}. This distinct goal is
accordingly pursued with a novel strategy for the identification of
classes that is reminiscent of the clustering technique used by Lacey
and Cole in an unrelated context \cite{lacey}. In the
following we shall try to propose a strategy able to perform an
optimal subdivision into classes of similarity and, for each of
these, provide the best representative. Two of the points of force of
such method are the absence of any subjectivity or human supervision
through the extensive use of optimal knowledge-based classification
criteria and also the fact that similarity classes are automatically
extracted and ranked according to their frequency of appearance in
natural proteins. This wealth of knowledge-based information provided
by the procedure allows to choose the representative set that best
matches one's needs. The clustering procedure we used to partition the
fragments in suitable similarity classes is conveniently illustrated
by the two-dimensional example of Fig. \ref{fig:cluster} where 1000
points have been assigned randomly to 4 distinct clusters with the
same radius but different size (i.e. number of members). Considerable
information about the clusters can be obtained by analyzing the
histogram of the distance between all pairs of members in the set. At
the simplest level the histogram analysis can reveal two distinct
scenarios: a) no clusters are present or b) there are clusters with
comparable size and degree of internal similarity. In the first case
the histogram distribution is expected to crowd around an average
value in a bell-shaped fashion. In the second case, two distinct peaks
should occur: one corresponding to the typical distance within
classes, the other centered around the (larger) average distance of
pairs of members from distinct classes. In the case of very few [many]
classes, the first [second] peak dominates.

The inset of Fig. \ref{fig:cluster} shows the pair-distance histogram
for the set of points in our example. It is evident that it consists
of two peaks: the first one extends till about the radius of clusters
while the location of the second peak coincides with the typical
cluster-cluster distance.

Our goal is to exploit the information obtained from the pair-distance
histogram to identify first how many different clusters there are and
secondly the optimal representative of each cluster. To do so we
follow the intuitive expectation that the best representative of each
cluster is the one closest to the cluster center. A deterministic way
to identify the center of homogeneous clusters, is to find the member
with the largest number of other points within a suitably chosen
similarity cutoff (we shall term this number ``proximity score'').
Indeed, points further from the center will have fewer
neighbors. This ``election'' mechanism is reliable for large and
homogeneous clusters.

Hence, we start by choosing the first representative of the set as the
one with the highest proximity score. This identifies simultaneously
both the largest cluster and its representative. Next, we remove the
representative and its cluster from the set and recalculate the
proximity score of the remaining points and again we select the member
with the highest score. As before we removed it and its cluster and
proceed in this iterative fashion until the set of surviving points is
exhausted.

When such scheme is applied to the set of Fig. \ref{fig:cluster} --
using a similarity cutoff equal to $R$=1 -- the optimal
representatives of the four clusters (marked with squares) are
immediately found and ranked according to their cluster size.

\subsection{Results}

We applied the same scheme to analyze our data bank of thousands of
protein fragments. This time, the points of the previous example are
replaced by the fragments themselves, while the notion of eucledian
distance between two points is substituted by the cRMS distance of two
fragments \cite{Unger}, $X$ and $Y$ of equal length, $N$,

\begin{equation}
\sigma(X,Y) =\sqrt{\sum_{k=1}^{N} \left| \vec{r}^{C_{\alpha}}_k (X) -
\vec{r}^{C_{\alpha}}_k(Y) \right|^2 \over N} \ .
\label{eqn:kabsch}
\end{equation}

\noindent This notion of distance is meaningful provided that $X$ and
$Y$ have been previously optimally superimposed with the standard
Kabsch procedure \cite{kabsch} . The calculation of the cRMS of each
distinct pair of fragments is the most computationally demanding step
since it requires an application of the Kabsch algorithm \cite{kabsch}
for each distinct pair of fragments (e.g. this translates in well over
ten million pairs of fragments for lengths of the order of 5).

\begin{figure}
\centerline{\psfig{figure=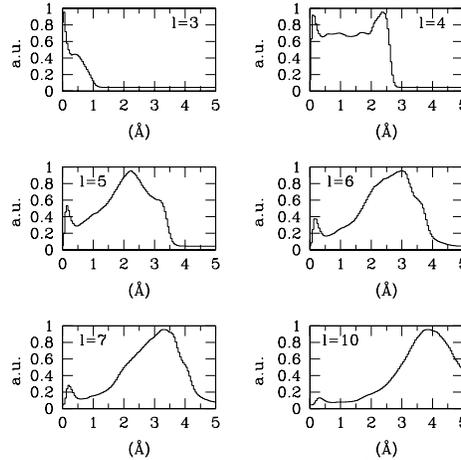,width=2.5in}}
\caption{Histogram of the distribution of distances between all pairs
of fragments of different length, $l$ extracted from the data bank of
Table \protect{\ref{tab:learn}} (the y-axis is in arbitrary units).}
\label{fig:histo}
\end{figure}

The histogram of all cRMS of pairs of fragments of lengths in the
range $3 \le l \le 10$ is given in Fig. \ref{fig:histo}. It can be
seen that all distributions show two distinct peaks, with the
exception of $l=3$, which appears to be exceptionally short, and hence
will be omitted from further analysis.

For the smaller lengths, the first peak collects a substantial amount
of ``hits'', proving that it is meaningful to assume the presence of
classes of similarity. It also appears that the height of the first
peak constantly decreases with increasing $l$. This confirms the
intuition that, by considering very large values of $l$ every fragment
will be a class for itself. Indeed, for lengths greater than 6, the
first peak is hardly discernible from the background. Hence, the mere
visual inspection of histogram distributions shows that it would not
be justifiable to force the introduction of classes of similarity for
lengths above 6. Nevertheless, we shall often present results also for
length 7 for the purpose of showing how several unrelated criteria
indicate such length as a border-case of viable oligons. An important
observation for our subsequent analysis is that the extension of the
first peak (the intra-cluster one) depends only weakly on $l$ and is
about 0.65 \AA. This provides an unbiased measure for the similarity
cutoff and hence we adopted it. The location of the second
``background'' peak in the histogram of Fig. \ref{fig:histo} gives an
estimate of the similarity between unrelated fragments and, hence,
crresponds to the cRMS deviation of a random pair of segments. This
random pair distance increases with the chain length, but it always
well above the value of 2 \AA, thus justifying {\em a posteriori}\ the
use of similarity cutoffs of the order of 1 \AA $\ $ considered in previous
studies \cite{Unger,a1}.

The advantage of the clustering scheme introduced and used here is
that, with modest computational effort (the cRMS distances need to be
computed once for all) one has simultaneously both the subdivision in
clusters and their optimal representatives. An extra payoff of this
approach over other clustering schemes is that the representatives are
singled out in order of importance. It is important to stress that
there is no stochastic element in the analysis since the assignement
of elements to clusters follows a ``greedy'' deterministic
approach. One particular instance where the suggested strategy may
fail, is when the ``fringes'' of distinct clusters overlap, that is
when an element falls in the similarity basin of more than one
representative. In this situation, more sophisticated clustering
techniques (such as those based on k-means analysis \cite{kmean})
ought to be adopted in place of the present one, in fact, the
iterative removal of assigned members would affect both the choice of
the representative and also its score.  Although we cannot rule out
the presence of fringe overlaps in our data bank, we can exclude it
has any substantial significance. Indeed, we have checked that the
typical cRMS of the extracted representatives matches the random pair
distance which, being much greater than 0.65 \AA, makes overlaps
highly improbable.

\begin{figure}
\centerline{\psfig{figure=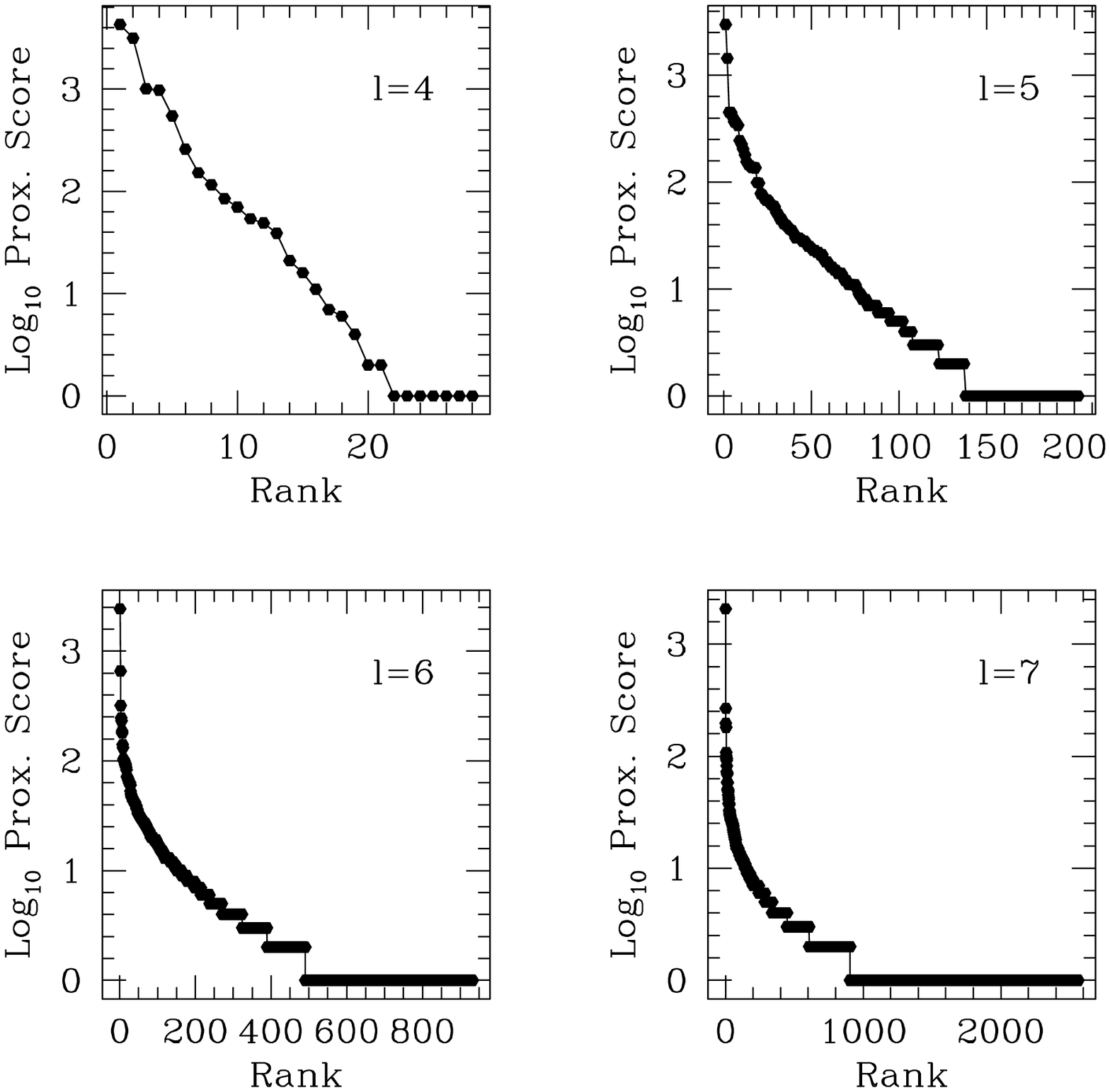,width=2.5in}}
\caption{Proximity score (in $\log_{10}$) versus ranking for the
representatives of the thousands of fragments of length $ 4 \le l \le 7$.}
\label{fig:rank}
\end{figure}

Our analysis identified only 28 representatives for length $l=4$, 202
for $l=5$, 932 for $l=6$ and 2561 for $l=7$. As we mentioned before,
the existence of a limited repertoire of local folds is a consequence
of the existence of a discrete number of degrees of freedom per amino acids,
as pointed out by the seminal studies of Levitt on $n$-state models
\cite{Levitt}. The results obtained here contain significantly more
knowledge-based information since, for instance, they also yield the
representation score of each representative. It appears that the
representation weight (i.e. proximity score) of the fragments
decreases very rapidly with the rank (see Fig. \ref{fig:rank} and
Table \ref{tab:winners}). This is an extremely important feature since
it indicates that one might discard the representatives with
negligible score and hence work with a subset of the whole data bank.
This issue is examined in the next section.

\begin{figure}
\centerline{\psfig{figure=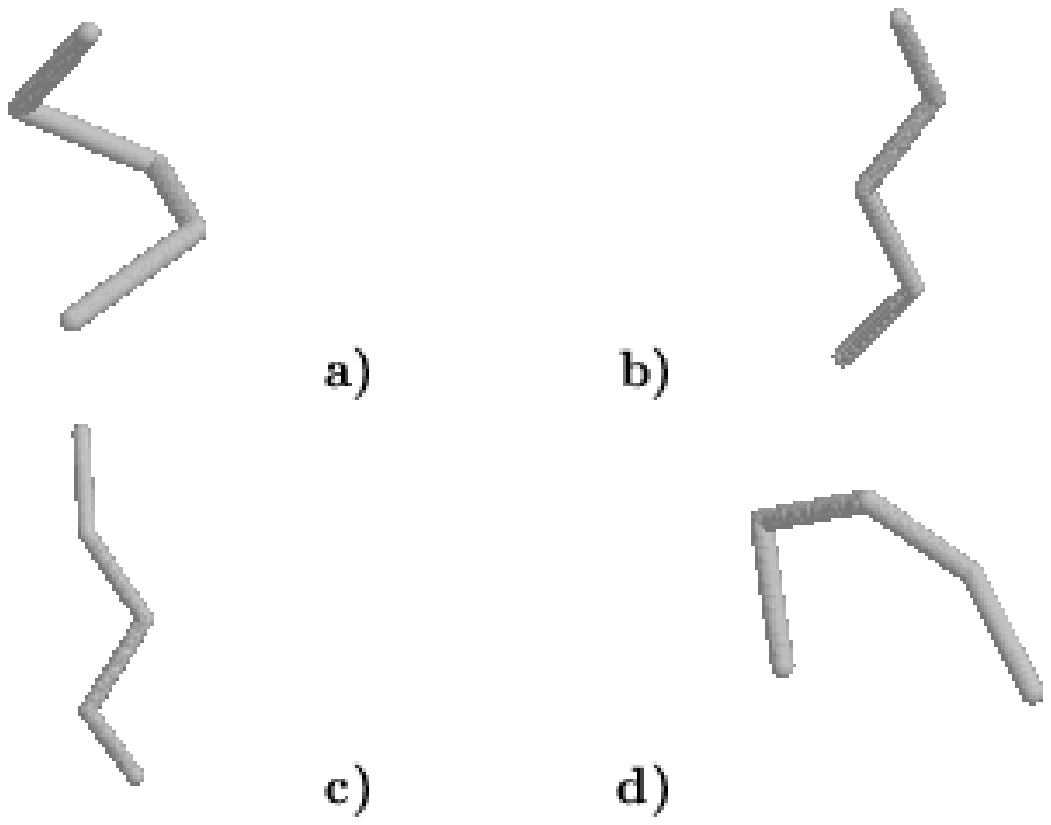,width=2.5in}}
\caption{The four oligons with the highest proximity score for $l=5$.}
\label{fig:best5}
\end{figure}

\begin{figure}
\centerline{\psfig{figure=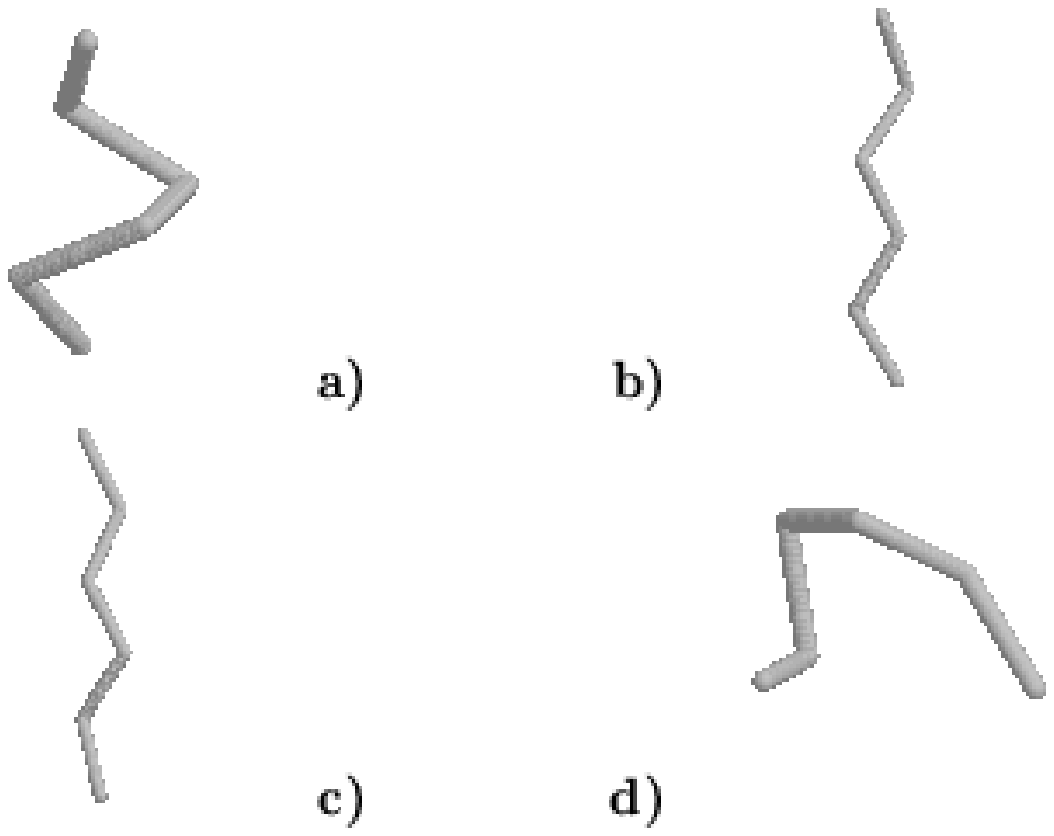,width=2.5in}}
\caption{The four oligons with the highest proximity score for $l=6$.}
\label{fig:best6}
\end{figure}

\begin{figure}
\centerline{\psfig{figure=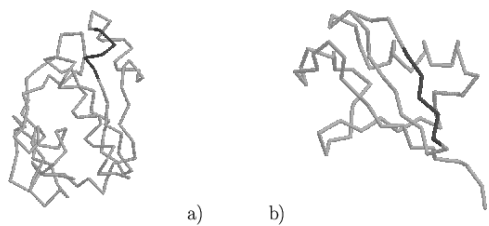,width=2.5in}}
\caption{The best two representative for $l=5$ shown in their native
protein environment.}
\label{fig:best1}
\end{figure}

One may expect that the best representatives should belong to the most
common structural motifs such as helices, strands or turns, whereas
the less frequent ones should correspond to the atypical parts of
proteins (structure exceptions). This expectation is confirmed by
inspection of the actual shape of the highest ranking fragments; the
consensus with the work of \cite{roomana,roomanb} and Unger
\cite{Unger} shows the reliability with which the main motifs can be
identified in different contexts or with different methods.  The first
four oligons for $l=5$ and $l=6$ are shown in Figs. \ref{fig:best5}
and \ref{fig:best6} (the structural data for the complete sets is
available at the URL given in the introduction). The native
environment of the first ten fragments of length 5 and 6 are given in
Table \ref{tab:winners}. For $l=5$ we have also shown the native
environments of the best representatives in Fig. \ref{fig:best1}. A
striking outcome of the clustering analysis is that the first 15
oligons of length 5 and 6 represent over 75 \% and 47 \%,
respectively, of the whole data-bank fragments! To the best of our
knowledge, this are the smallest sets of representative fragments able
to cover most local structural instances with an uncertainty comparable
with the best experimental resolution.

\section{Discussion}
\label{sec:disc}

\subsection{Analysis of the clustering procedure}

Before testing the goodness of the representative fragments it is
necessary to validate the clustering procedure and ensure that the
results are robust and not too dependent on the details of the data
bank. We carried out a first check by studying how the outcome of the
clustering scheme is affected by the size of our data bank. To be
precise, this test goes beyond the mere validation of the oligon
extraction scheme, since it also constitutes a check of the
applicability of any clustering scheme to protein fragments. To
proceed in an unbiased way we randomized the order of fragments in the
data bank, so to cancel correlations of consecutive (overlapping)
oligons, and extracted the representatives for an increasing number of
fragments taken from the top of the randomized list.  A careful
analysis of the data has revealed that, for any length, $l$, the
number of trivial representatives, i.e. those that, having score equal
to 1, represent only themselves, grows linearly with the size of the
data bank. The preportion of trivial representatives is about 0.5 \%,
2\% of the whole population for lengths 5, 6. For length greater than
6 the proportion of trivial representatives is considerable (being
greater than 10 \%). On the other hand, for $ 4 \le l \le 6$ the
number of non-trivial representatives shows very little increase with
the size of the data bank and can be considered constant for all
practical purposes. This provides a solid {\em a posteriori}
confirmation that the data-bank is of sufficiently large size.  Of
course, the number of representatives and their growth with data-bank
size depends on the particular choice of similarity cutoff (the
smaller the value, the larger the number of classes). In this
particular study the choice of the cutoff was dictated by the
properties of the very same data to be clustered. Nevertheless, the
use of physically viable cutoffs lead, invariably, to the
identification of the same high-ranking clusters and, correspondingly,
almost identical representatives. This could be expected {\em a
priori}\ since identifying the most common local folds should be
independent, to a large extent, on the details of the clustering
procedure. As explained in the next sectiom, we tried to build on this
robust result and concentrate only on the top representatives.

\subsection{Reducing the representative sets}

 Since the trivial oligons mentioned in Sec. III A represent only
themselves, one may wonder if they can be dropped from the set and
still be able to represent well the majority of native structures. In
this subsection we considered this problem and try to quantify the
attainable accuracy in representation when a subset of the
representatives is used. For a preassigned number, $m$, of
representatives to be used, the optimal accuracy is obtained when the
$m$ highest ranking fragments are taken. Hence, our extraction scheme
is particularly convenient for this type of study since it yields the
representative fragment ranked according to their proximity score

In this framework, we measure the accuracy of representation by the
amount of local structural deformation required to bring all fragments
of the native structure within the proximity basin of any of the
reduced oligons (while in Sect. IIID we shall fit several proteins
with the oligons). This is a sort of measure of the ``completeness''
of the set of oligons: if the used set of oligons represented all
possible instances of protein fragments, no deformation would be
required. On the contrary, the poorer the set of representatives, the
larger is the deformation required to bring the original fragments in
the proximity basin of one oligon. To do so, we use a stochastic Monte
Carlo dynamics on the backbone (described in Appendix A) to minimise
the following quantity:

\begin{equation}
S \equiv \sum_{i=0}^{L/l-1} (\sigma(B_i,\omega_i) - R)^2 \cdot
\theta[\sigma(B_i,\omega_i) - R]
\label{eqn:fit}
\end{equation}

\noindent where $L$ is the length of the protein, $\sigma$ is the cRMS
distance of eqn. \ref{eqn:kabsch}, $B_i$ is the $i$th backbone
fragment of length $l$, $\omega$ is its closest oligon, R is the
similarity distance (0.65 \AA) and $\theta$ is the usual step
function.

\begin{figure}
\centerline{\psfig{figure=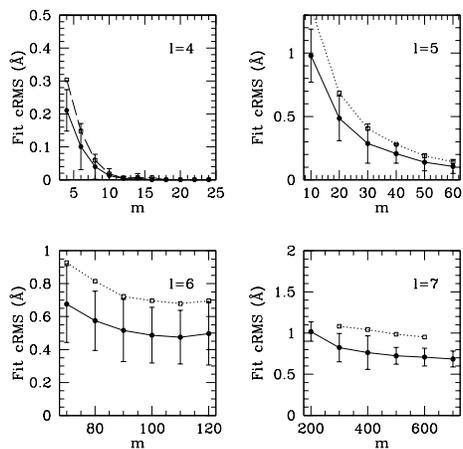,width=2.5in}}
\caption{When a subset of $m$ ranked oligons in used, not all
arbitrary fragments of protein backbones can be represented within
0.65 \AA. In this plot we show (solid lines) how much, on average it
was necessary to distort the ten proteins in the test set so that each
of their fragments fell within the proximity basin of the first $m$
ranked oligons. The dotted lines show the average deviations of the
fitted and native $C_\beta$ positions computed for the test proteins.}
\label{fig:fit}
\end{figure}

\noindent By using the stochastic dynamics, the starting structure is
deformed until all fragments are within the preassigned distance R
from one of the oligons. When this happens, the score function, $S$,
is exactly zero and the dynamics is stopped. By measuring how far (in
terms of cRMS) the backbone has moved from its original position we
can judge whether the achievable quality of representation is
acceptable. We carried out this scheme by using only the first few
representatives (for each length $ 4 \le l \le 7$ ) and then increased
their number progressively (always choosing the highest ranking
ones). The cRMS as a function of the number of representatives is
shown in Figs. \ref{fig:fit}. For $ 4 \le l \le 6$ only a fraction of
the collected oligons are necessary to fit the 10 test structures
within 0.65 \AA\ and with no need to distort them. Even for $l=6$ with
only 100 representatives any protein backbone can be fitted at the
price of minute distortions (less than 0.5 \AA\ ), that are finer than
the typical experimental structural resolution.

It is important to point out the the low global cRMS values given in
Fig. \ref{fig:fit} do not hide exceedingly large local distortions
averaged with many other smaller local deviations. Indeed, the
deviations appear to be homogeneous along the chain; the worst local
distance of fitted fragments from the native positions never exceeds
twice the global averaged value (data not shown). To be more precise
in assessing the presence and effects of unphysical local deviations
we calculated the displacementes
of $C_\beta$ positions in fitted backbones from the native
one. Indeed, $C_\beta$'s discrepancies are good indicators of local
variations in the dihedral angles between virtual $C_\alpha$
bonds. $C_\beta$ positions were recovered from $C_\alpha$ coordinates
$\{\vec{r}^\alpha_i\}$ through the standard geometric constrained
construction \cite{Levitt}:

\begin{equation}
\vec{r}_i^{\beta} = \vec{r}^\alpha_i + d_0 \left( \hat{a} \cdot
\cos{(\theta)} + \hat{b} \cdot \sin{(\theta)} \right)
\end{equation} 

\noindent where:

\begin{equation}
\hat{a}= \frac{\hat{s}_{i,i-1}+\hat{s}_{i,i+1}}
{\left|\hat{s}_{i,i-1}+\hat{s}_{i,i+1}\right|} \ \ \
\hat{b}= \frac{\hat{s}_{i,i-1} \land \hat{s}_{i,i+1}}
{\left|\hat{s}_{i,i-1}\land \hat{s}_{i,i+1}\right|} 
\end{equation}

\noindent and:

\begin{equation}
\hat{s}_{i,j}=\vec{r}^\alpha_i-\vec{r}^\alpha_j \ .
\end{equation}

In the previous formulae $d_0=3$ \AA\ is the distance of the
$C_{\beta}$ atoms from the corresponding $C_{\alpha}$ atom and
$\theta$ is the out-of-plane angle optimally set to $37.6^{0}$. The
$C_{\beta}$ positions are very sensitive to the local position of the
$C_{\alpha}$ because a wrong (even by a small amount) choice of the
angle between the $\hat{s}_i$ can heavily affect its position,
e.g. shift it to the wrong side of the chain.

We considered some of the proteins in the test set previously fitted
with a subset of the representative oligons. For these we constructed
the $C_\beta$ positions and calculated the deviations of the latter
from those in the native configurations. the data are shown with
dotted lines in Fig. \ref{fig:fit} and highlight how the discrepancy
is very small and follows the trend of the cRMS for $C_\alpha$
atoms. This shows that the local distortions are really tiny even when
100 of the over 600 oligons of length 6 are used. As usual, an
atypical behaviour is seen for length 7, for which, even using
hundreds of fragments, a much larger discrepancy is observable.

\subsection{Optimal length of representative oligons}

Each of the sets of representative fragments of length $4 \le l \le 6$
are optimal by construction and all of them satisfy the rigid tests
carried out so far. The goal we pose here is to decide which length is
the best. The answer is certainly not unique, since different criteria
for optimality can be used \cite{Unger,roomana,roomanb}. For example,
if one is interested in having the smallest possible set of
representatives, then small values of $l$ are to be preferred. On the
other hand, if one is mainly interested in having the least number of
conformational degrees of freedom per residue then $l$ should be
chosen as large as possible. Both approaches can be legitimate in
appropriate contexts.  From a general point of view, however, using
very short fragments defeats the purpose of this study - that is to
capture structural correlations. On the other hand, excessively large
values of $l$ are more difficult to handle and uninteresting since
clusters will typically be sparsely populated (over specialised
case). Here we examine the main properties of representative oligons
that can be conveniently exploited in different contexts. We begin by
discussing how well oligons of different length represent secondary
motifs \cite{roomana,roomanb,a1,a7,a8}.  The latter are indeed the
distinctive feature of proteins (as opposed to random heteropolymers
\cite{pauling,onuchic,rose,aurora2,gregoret,M00} and have several
consequences on biophysical properties, such as speeding up the
folding process or providing maximum kinetic accessibility to the
native state \cite{Micheletti99b}.

Alpha-helices seem to be fairly easy to represent. In fact, for all
cases $l=4,5,6,7$ a single representative (namely the highest-scoring
one) is sufficient to represent virtually all instances of
helices. The situation is different for $\beta$-strands, due to the
different environment in which they can be found (parallel or
anti-parallel, bent $\beta$-barrels, Greek-key motifs etc). This
variability implies that more than one representative for $\beta$
motifs is found (although not with the same proximity score) .
Examples are shown in Figs. \ref{fig:best5} and \ref{fig:best6}. This
proliferation effect is more dramatic for longer fragments,
consistently with the findings of Prestrelsky et
al. \cite{a1}. Indeed, for $l=7$, each of the distinct $\beta$ classes
appears severely depleted, containing typically less than 100 elements
which is a small fraction of the score of the helical one (2070).

For all values of $l$, however, the largest number of representatives
is covered by segments representing loop regions. These results are
particularly relevant for modeling/characterizing regions of high
variability, but our main focus is on the possibility to represent
synthetically, though accurately, recurrent oligons. Within such
minimalistic approaches, the choice of representatives of length 5
seems to be the best one, since it captures non-trivial correlations
while using essentially a single representative for $\alpha$ and
$\beta$ instances.

\subsection{Fitting proteins with oligons}

\begin{figure}
\centerline{\psfig{figure=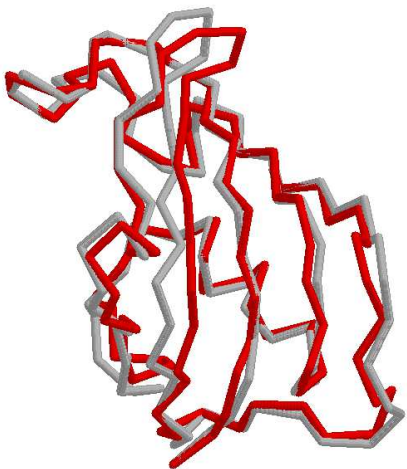,width=2.5in}}
\caption{Illustration of the rigid fit procedure. The crystallographic
structure of protein 1fkb (dark backbone) has been fitted by using the
limited set of the first 40 oligons of length 5 (lighter backbone).}
\label{fig:vhhfit}
\end{figure}

Another criterion for selecting the most suitable length is how well
can we reproduce a given protein by ``gluing'' rigidly together only
the representative oligons? The purpose of such question is to
investigate the benefit of employing oligons in folding contexts. A
simple and powerful way to speed up the numerical simulations of
folding would be to consider structures made only by ``gluing''
suitably chosen representative oligons. In such framework the only
degrees of freedom that one has to contend with are: 1) which oligon
to use and 2) how to connect successive oligons. This is a severe
reduction of the traditional continuous/discrete degrees of freedom
per amino acid adopted in ordinary Monte Carlo or Molecular Dynamics
schemes. The feasibility of such scheme depends first of all on the
possibility to reproduce sufficiently well any given native structure
by joining rigidly the oligons. We checked this by following a
stochastic process to find both the best oligons to be used locally
and also their best relative orientations. This was almost a
worst-case scenario due to the independence of the test set from that
of Table \ref{tab:learn}. The optimal fit was accomplished by
progressively distorting the native structure with the local Monte
Carlo moves described in Appendix A.  The "energy-like" cost function
had the same form of (\ref{eqn:fit}), but where $R$ is set to an
arbitrary small positive quantity, $10^{-3}$ in our case.  Again, we
carried out the stochastic dynamics (proceedings through very tiny
local deformations) until the cost function was reduced to zero.  This
signalling that each protein fragment had been optimally collapsed on
an oligon. It can be anticipated that, due to the propagation of
misfits, the cRMS with respect to the native protein would be rather
larger than the similarity cutoff of 0.65 \AA. Moreover, it may be
expected that smaller oligons may lead to smaller cRMS since they
might provide more flexibility in ``tiling'' target structures.
Surprisingly, this is not the case, as visible in Table
\ref{tab:rigfit}, where we summarised the global cRMS deviations for
rigidly fitting the 10 proteins in the test set.  Remarkably, the
overall cRMS is always very close to 1 \AA\ such cRMS deviations of
the native and fitted protein can be appreciated visually in
Fig. \ref{fig:vhhfit}. We explain the little dependence of cRMS fits
on oligon lengths with the observation that, irrespective of the
oligon length, each residue in native conformations is typically 0.5
\AA\ away from the corresponding position in the best-matching
oligon. This little sensitivity on $l$ is, in turn, reflected on the
overall cRMS of the rigid fit. The fit discrepancy is not only
independent of the length but also fully compatible with
state-of-the-art experimental resolution of crystallographic
structures. For these reasons one may adopt oligons of the longest
possible lengths if the primary interest is capturing the longest
possible structural correlations. This would suggest to consider
lengths equal to 6. Our fit scheme has considerable advantages over
previous ones where representatives obtained with different techniques
were employed. For example, in their classic paper, Unger {\it et al.}
\cite{Unger} used a molecular best fit procedure that yielded cRMS of
over 7 \AA\ when hexamers were used to fit peptides of over 70
residues. The dramatic improvement of the results in Table
\ref{tab:rigfit} confirms the validity and reliability of both the
clustering method and of the extracted set of oligons. Indeed, the low
values of cRMS fit support the expectation that the extracted oligons
can be successfully used to speed up folding attempts. Preliminary
tests in this direction have been carried out in folding contexts
where perfectly smooth folding funnels \cite{Go} lead to known
crystallographic structures. Such studies originally undertaken to
elucidate global aspects of the folding process have recently been the
key to predict and describe the influence of topological protein
properties on folding nuclei and/or thermodynamical folding stages
\cite{Micheletti99b}. By employing oligons of length 5 we were able to
speed up the collection of folding data by several factors
\cite{prep}.

\subsection{Correlation between oligons and amino acid sequences}

We devote the final part of this section to elucidate the possibility
of finding correlations between oligons and amino-acids sequences.  In
general, it is well-known that there is preference for definite sets
of amino acids to occupy or avoid specific structural motifs
\cite{BT91,C92,rose99}.  Here we examine the extent to which such
propensities are reflected in the oligons and the clusters they
represent.  Highlighting connections between sequences and oligons has
a twofold purpose: a clear preference of an amino-acid sequence to be
mounted in a specific oligon can be useful exploited in folding
predictions, whereas design attempts can be greatly aided by
discovering that some oligons preferably house very few sequences.

The connection between sequence-structures connections have been
heavily investigated, with fair success, for a variety of fragment
lengths and amino-acid sequences. It is important to examine the issue
also in the present context since the emergence of clear correlations
between sequences and oligons could be an additional aid in reducing
the computational complexity of folding and/or design.

For sake of simplicity we consider in this section only the case
$l=5$ and we considered the best $40$ oligons of that length.
 We start by introducing a suitable classification of the 20
types of amino acids. This is essential to proceed, since otherwise
the shear number of the possible sequences, $20^5 \approx$ 3 million,
would make it impossible to gather sufficient statistics for all
quintuplets. The classification scheme we introduce here is based on
some general results for chemical affinities
\cite{classes,mayo,BT91,Micheletti98b,huang,chan} and some empiric
attempts.  According to it we subdivide the residues in four distinct
classes.

In the first we place Gly, in the second Pro, in the third the
hydrophobic (H) aminoacids (Ala, Val, Leu, Ile, Cys, Met, Phe, Tyr,
Trp) and finally in the fourth the polar (P) ones (Hys, Ser, Thr, Lys,
Arg, Asp, Asn, Gln, Glu). With this subdivision we keep separate the
amino acids (Gly and Pro) that can attain atypical
conformations/chiralities \cite{rose99} (and hence may act as helix
breakers etc.).  It is also wise to keep in separate families
hydrophobic and polar aminoacids, since they can alternate regularly
in secondary motifs partly exposed to the solvent \cite{BT91}. Within
this framework we could obtain in principle up to $4^5=1024$ distinct
pentamer sequences (we always consider our pentamers as ``directed''
in that the C and N termini are not exchangeable). It turns out that,
due to chemical and steric constraint, not all pentamer sequences are
observed in nature, and hence in our data-bank.

\begin{figure}
\centerline{\psfig{figure=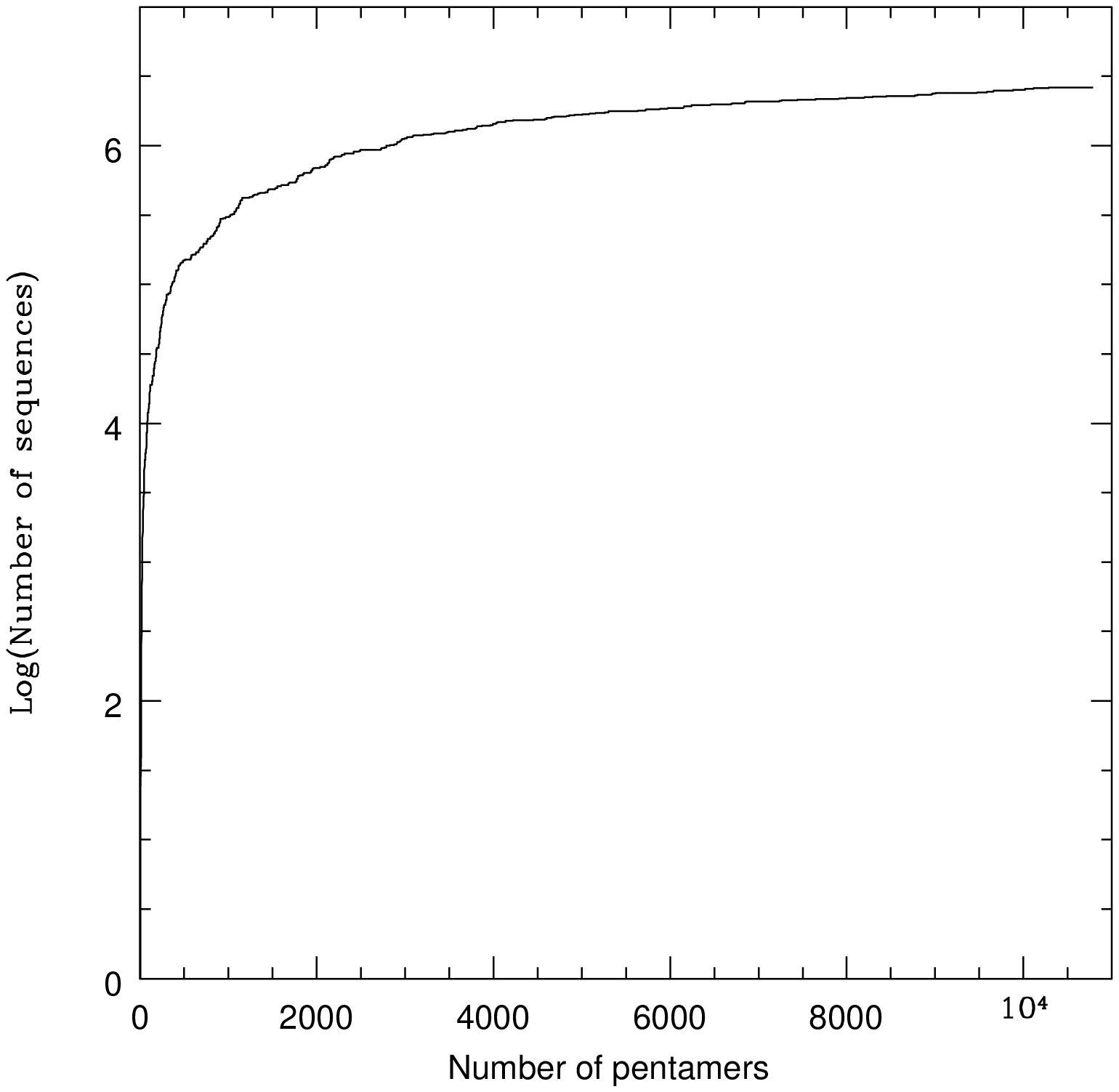,width=2.5in}}
\caption{Number of emerging sequences (in natural-logarithmic scale) as a
function of the considered fragments.}
\label{fig:log}
\end{figure}

To perform our analysis we considered all the proteins (75) appearing
in Table  \ref{tab:learn} . We partitioned then
in overlapping fragments of length 5 ending up with 10786 pentamers.
The size of this data-bank was sufficient to provide excellent
coverage of all possible pentamer sequences. This is evident from the
plot of Fig. \ref{fig:log} which shows how the number of distinct
pentamer sequences grows with the data-bank size.

The asymptotic number of distinct sequences we obtained from the near
six thousands instances was 614, about half of all possible ones.

To match the 614 sequences to the 40 oligons of length 5 we re-applied
the clustering procedure: to each of the oligons we assign not only
its native sequence but also those of each member in the cluster it
represents. All this information can be conveniently stored in a score
matrix $z(i,j)$ whose entries correspond to the number of times that
the $j$th sequence has been assigned to the $i$th oligon (hence $z$ is
a 40x614 matrix);

\begin{figure}
\centerline{\psfig{figure=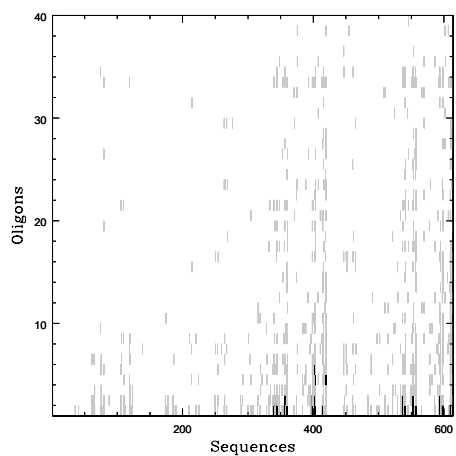,width=2.5in}}
\caption{Two dimensional representation of the score matrix. In the
x-axis the 614 sequences are labelled according to a conventional
order. In the y-axis the best 40 oligons of length 5 are labelled
according to their proximity score. The intensity of the colour is
related to the values of the entries: the blank areas denote entries
in the range 0-2, grey for the range 3-25, black for entries greater
than 25. }
\label{fig:bestiale}
\end{figure}

A two-dimensional representation of the score matrix is plotted in
Fig. \ref{fig:bestiale} where the dark boxes correspond to entries
above 25, the grey ones to entries between 3 and 25 and the blank ones
to  entries below 3 . The figure shows that $z$ is a sparse matrix,
since only few entries have a significative entry (bigger than 3). 
This supporting the
conjecture of strong correlations between oligons and sequences.

\begin{figure}
\centerline{\psfig{figure=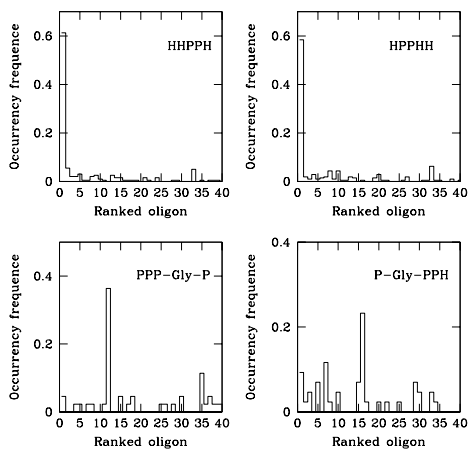,width=2.5in}}
\caption{Histograms showing the relative frequency with which four sequences
occupy ranked oligons. The oligons are ranked
according to their proximity score.}
\label{fig:quattros}
\end{figure}

The last observation can be turned into a more quantitative statement
by examining the behaviour of definite oligons and/or pentamer
sequences. The natural candidates to focus on are the 135 sequences
that appear more that 20 times, and hence allow a statistically sound
analysis. For each one of these sequences we examined the relative
frequency with which they occupy a given oligon.  Typical results are
given as histograms in Fig. \ref{fig:quattros}.

\begin{figure}
\centerline{\psfig{figure=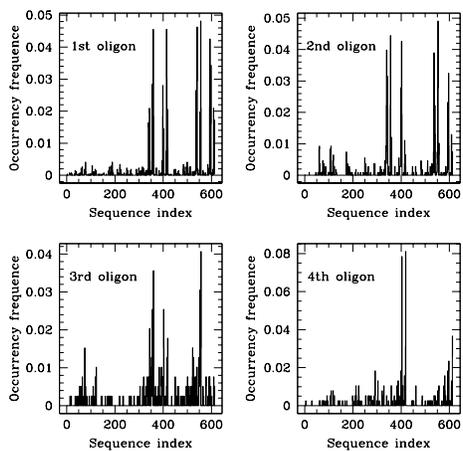,width=2.5in}}
\caption{Histograms showing the relative frequency with which the
first four oligons (ranked according to their proximity score) house
different sequences.  For this plot the same sequence-indexing of
Fig. 10 is used}
\label{fig:quattrofol}
\end{figure}

It appears that sequences do not occupy many oligons; in fact, less
than 18 oligons are occupied, on average, by the 135 sequences ( and
over 70 \% of the entries is covered by six oligons). It is worth
underlining how this is not an average effect reflecting the relative
magnitude of the proximity scores of the oligons.  To show this one
can establish a reference threshold corresponding to the number of
expected hits if sequences are distributed uniformly over all
fragments. Thus, for a given oligon, the threshold is simply the ratio
between its proximity score and the total number of fragments used to
calculate this score. It was found that in 103 cases out of 135 ( 77
\% ) the sequences select their preferred oligon with a percentage
significantly higher (in excess of $20 \%$ than the trivial
threshold).  Although it is clear that any given sequence is
compatible only with few oligons, the converse is not true. This
interesting asymmetry between sequence and structure has deep roots,
as first shown by Anfinsen \cite{Anfinsen73}, who pointed out that a
protein sequence uniquely identifies its structure, while several
different sequences can admit (almost) the same structure as their
native states.  This aspect is strikingly evident when plots analogous
to those of Fig. \ref{fig:quattros} (by interchanging the role of
sequences and oligons) are made. In Fig. \ref{fig:quattrofol} the
occurrence frequency histograms for the first (ranked according to the
proximity score) four oligons are plotted. In these histograms for
each sequence (listed in ordinate according to a convenient scheme)
the percentage of occurrence for the given oligon is represented.

It is clear that, unlike the case for pentamer sequences, there is not
a preference for a given oligon to be occupied by few sequences, so
that the benefits of these correlations studies for design schemes is
not as dramatic as could be for folding simulations.

As a final test we verify whether it is possible to define selection
rules for locating amino acids in well-defined oligon positions,
e.g. to pinpoint particular points where it is unlikely that some
class of amino acid could appear. The existence of such forbidden
points could be, again, a useful source of information for folding and
design. Due to the non-homogeneous population of the amino acids
classes we adopted, we expect to extract information only for the
first two classes, namely Gly and Pro. For any oligon we considered
all the related sequences and we monitored, site by site, the
occurrence frequency of each class. If for a given site and class this
frequency is below the threshold of 0.5\% we consider the event
unprobable (and hence significant in the present context). In Table
\ref{tab:forbidden} we list 19 of these events which take place in the
highest-ranking oligons. The most significant instances all refer to
Pro ``class''.  The final test we have summarised shows that a
combination of the structural reduction in oligons and associated
correlations with local sequence propensities can be turned into a
powerful tool in aiding folding and design. This hope is corroborated
by the recent successes of structural prediction schemes based on
local sequence propensities \cite{baker,baker2}.

\section{Conclusions}
\label{sec:con}

The starting point of this work was the conclusion of recent previous
studies that there exist recurrent local motifs in natural proteins
\cite{Unger,roomana,roomanb,a1}.  We introduce novel and fully
automated criteria for an optimal partitioning of a complete data-bank
of fragments taken from non-redundant proteins into classes of
similarity.

We exploit the intrinsic information in the data-bank to identify the
classes with the least bias or human supervision. Our goal was to show
that such scheme succeeds in conciling two competing aspects of
protein modeling: accuracy and synthetic modeling \cite{Levitt}.
 
In fact, on one hand this method is shown to provide the most economic
subdivision in classes (the number of which is not set a priori). On
the other hand, the optimally extracted representatives from each
class are shown to be sufficient to represent and fit virtually all
protein structures with an uncertainty of 1 \AA (rigid fitting) or 0.5
\AA, when only local similarity within the proximity basin is
required. We also considered several possible lengths for oligons and
examined their suitability in different modelling contexts. It turns
out that $l=5$ is the most suitable when the smallest representative
set is needed, while $l=6$ is best when it is necessary to capture the
longest possible correlations. Lengths smaller than 5 or longer than 7
appear to be far from optimality.

\section{Acknowledgments}

We acknowledge support from the Theoretical and Biophysical sections
of the INFM. We are indebt to the  Italian
Research Council  for the financial support of the advanced research project
``Statistical mechanics of proteins and random heteropolymers''.

\appendix \section{Monte Carlo dynamics}

In this Appendix we present a summary of the stochastic approach that
we used for the dynamics of the protein backbones. As mentioned in the
text we used the Monte Carlo dynamics for progressive distortion of
native protein backbones in order to fit them locally by using a
restricted set of representative fragments (see Section
\ref{sec:disc}) to provide the best protein fit by using exactly the
representatives. In the spirit of standard dynamical approaches for
three-dimensional structures \cite{Gerroff,Sokal,kgs} each time we
propose a Monte Carlo move we distort the structure by performing
either local or global rearrangements. Local moves are single-bead or
crankshaft, as explained below, while pivot rotations were employed
for global ones.

\begin{figure}
\centerline{\psfig{figure=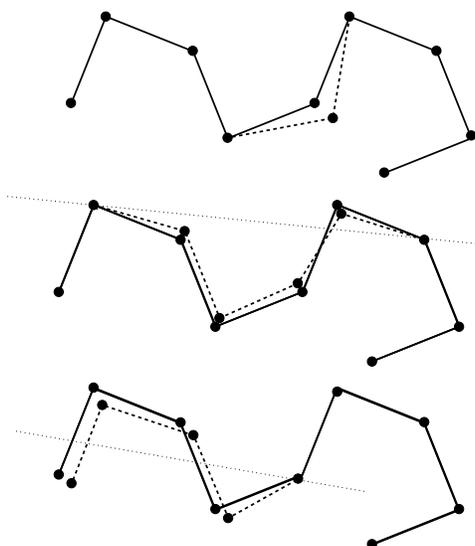,width=2.5in}}
\caption{Monte-Carlo moves: (top) a single bead is moved; (middle) a
set of amino-acids is moved by rotating a portion of the protein
around a fixed axis; (bottom) a set of amino-acids is moved by
pivotting part of the protein around a fixed point.}
\label{fig:mcmoves}
\end{figure}

In the following we will use the ordinary Cartesian triplet $(x,y,z)$
to indicate the co-ordinates of $C_\alpha$ atoms. Subscripts will
denote the amino acid position along the sequence. The three types of
moves are as follows:

\begin{enumerate}

\item {\it Single $C_{\alpha}$ move. } A random site $i$ of the
protein chain is chosen and its old coordinates are replaced by new
ones $(x_i^{'},y_i^{'},z_i^{'})$ defined as:

\begin{equation}
x_i^{'} = x_i + \eta_1 \Delta l \ \ \
y_i^{'} = y_i + \eta_2 \Delta l \ \ \
z_i^{'} = z_i + \eta_3 \Delta l 
\end{equation}

\noindent where $(\eta_1,\eta_2,\eta_3)$ are three independent random
numbers in the interval $(-1,1)$ and $\Delta l$ is a distance that we
fixed (see discussion below) equal to 1 \AA\ (top panel of
Fig. \ref{fig:mcmoves}).

\item {\it Crankshaft move}. Two protein sites $i$ and $j$ with
sequence separation at most $10$ are chosen. Then the all the sites
between $i$ and $j$ are rotated around the axis going through $i$ and
$j$ by a random angle in the range $-\frac{\pi}{10} \leq \theta \leq
\frac{\pi}{10}$; (middle panel of Fig. \ref{fig:mcmoves}).

\item {\it Pivot move}. A random site $i$ and a random axis passing
through it are chosen. All the sites from $i+1$ to the end are then
rotated around the axis by an angle in the range $(-\frac{\pi}{10}
,\frac{\pi}{10})$; (bottom panel of Fig. \ref{fig:mcmoves}).

\end{enumerate}

The new configuration generated by applying one of these moves (chosen
with equal weight) is first examined to make sure that it does not
violate basic geometrical constraints obeyed by natural proteins, namely:

\begin{enumerate}
\item the distance between two consecutive $C_{\alpha}$ atoms
(measured in \AA\ ) must remain in the range $(3.7,3.9)$ and
\item the distance between two non consecutive $C_{\alpha}$ atoms must
be greater than $4 \AA\ $.
\end{enumerate}

\noindent If these conditions are not fulfilled, then a new move is
attempted. When the new configuration has passed the geometrical test
then is accepted/rejected through the classic Metropolis rule.

\newpage

\begin{table}
\begin{tabular}{|r|r|r|c|}\hline
Name & Length & Scop code & Family \\\hline \hline
1vii & 36 & 1001014001001 & 001 \\
1pru & 56 & 1001030001003 & 001 \\
1fxd & 58 & 1004033001001 & 001 \\
1igd & 61 & 1004012001001 & 001 \\
1orc & 64 & 1001030001002 & 005 \\
1sap & 66 & 1004009001001 & 002 \\
1mit & 69 & 1004022001001 & 003 \\
1ail & 70 & 1001015001001 & 001 \\
1utg & 70 & 1001072001001 & 001 \\
1hoe & 74 & 1002004001001 & 001 \\
1kjs & 74 & 1001040001001 & 001 \\
1hyp & 75 & 1001042001001 & 001 \\
1fow & 76 & 1001004004001 & 001 \\
1tif & 76 & 1004012006001 & 001 \\
1tnt & 76 & 1001006001001 & 001 \\
1ubi & 76 & 1004012002001 & 001 \\
1acp & 77 & 1001026001001 & 001 \\
1vcc & 77 & 1004067001001 & 001 \\
1coo & 81 & 1001032001001 & 001 \\
1cei & 85 & 1001026002001 & 001 \\
1opd & 85 & 1004052001001 & 003 \\
1fna & 91 & 1002001002001 & 002 \\
1pdr & 96 & 1002023001001 & 001 \\
1beo & 98 & 1001096001001 & 001 \\
1tul & 102 & 1002060004001 & 001 \\
1aac & 105 & 1002005001001 & 001 \\
1erv & 105 & 1003033001001 & 004 \\
1jpc & 108 & 1002054001001 & 001 \\
1kum & 108 & 1002003001001 & 005 \\
1rro & 108 & 1001034001004 & 001 \\
1poa & 118 & 1001095001002 & 001 \\
1mai & 119 & 1002037001001 & 001 \\
1bfg & 126 & 1002028001001 & 001 \\
1pdo & 129 & 1003040001001 & 001 \\
1ifc & 131 & 1002041001002 & 002 \\
1lis & 131 & 1001017001001 & 001 \\
1kuh & 132 & 1004050001001 & 001 \\
1cof & 135 & 1004060001002 & 001 \\
1rsy & 135 & 1002006001002 & 001 \\
1lcl & 141 & 1002019001003 & 004 \\
1pkp & 145 & 1004011001001 & 002 \\
1lba & 146 & 1004064001001 & 001 \\
1vsd & 146 & 1003041003002 & 001 \\
1npk & 150 & 1004033006001 & 002 \\
1vhh & 157 & 1004034001002 & 001 \\
1gpr & 158 & 1002059003001 & 001 \\
1ra9 & 159 & 1003053001001 & 001 \\
119l & 162 & 1004002001003 & 001 \\
1sfe & 165 & 1001004002001 & 001 \\
1amm & 174 & 1002009001001 & 001 \\
1ido & 184 & 1003045001001 & 002 \\
153l & 185 & 1004002001004 & 001 \\
1knb & 186 & 1002016001001 & 001 \\
1kid & 193 & 1003005003001 & 001 \\
1cex & 197 & 1003013007001 & 001 \\
1chd & 198 & 1003027001001 & 001 \\
1fua & 206 & 1003055001001 & 001 \\
1thv & 207 & 1002018001001 & 001 \\
1ah6 & 213 & 1004068001001 & 001 \\
1lbu & 214 & 1001019001001 & 001 \\
1gpc & 218 & 1002026004007 & 003 \\
1akz & 223 & 1003011001001 & 001 \\
1dad & 224 & 1003025001005 & 001 \\
1cby & 227 & 1004058001001 & 001 \\
1aol & 228 & 1002015001001 & 001 \\
1lbd & 238 & 1001087001001 & 001 \\
1mrj & 247 & 1004094001001 & 001 \\
1plq & 258 & 1004076001002 & 001 \\
1arb & 263 & 1002031001001 & 001 \\
1ako & 268 & 1004086001001 & 001 \\
1tml & 286 & 1003002001001 & 001 \\
1han & 287 & 1004020001003 & 002 \\
1nar & 289 & 1003001001005 & 002 \\
1amp & 291 & 1003052003004 & 001 \\
1ctt & 294 & 1003075001001 & 001 \\
\end{tabular}
\caption{Non redundant proteins used to extract the oligons. 
The reported length is the one actually used in this work.}
\label{tab:learn}
\end{table}

\begin{table}
\begin{tabular}{|r||r|r|r||r|r|r|}\hline
&\multicolumn{3}{c||}{l=5}&\multicolumn{3}{c|}{l=6}\\\hline 
 Rank & Score & Parent & Location & Score & Parent 
& Location \\\hline \hline
1 & 2991 &  1mai &  81 - 85  & 2429 & 1orc &  25 - 30 \\
2 & 1442 &  1ubi &  10 - 14  &  658 & 1aac &  41 - 46\\
3 &  451 &  1amm & 167 - 17 &  319 & 1plq &  24 - 29\\
4 &  449 &  1akz &  17 - 21  &  246 & 1cex &  74 - 79\\
5 &  411 &  1ah6 & 208 - 21 &  231 & 1fna &  60 - 65\\
6 &  366 &  1ctt & 225 - 22 &  187 & 1sfe & 100 - 105\\
7 &  357 &  1cex &  94 - 98  &  179 & 1lis & 117 - 122\\
8 &  340 &  1akz &  15 - 19  &  141 & 1rsy & 128 - 133\\
9 &  245 &  1npk & 138 - 14 &  132 & 1cex &  93 - 98\\
10 &  227 &  1akz &  56 - 60 &  104 & 1aac & 39 -  44\\ \hline \hline
\end{tabular}
\caption{The first ten oligons for $l=5$ and $l=6$ ranked by proximity
score. In the third column the PDB code of the protein from which they
have been extracted and in the fourth column their position along the
backbone chain (amino acids are indexed starting from the beginning of
the pdb file, regardless of the numeration in the pdb file itself).}
\label{tab:winners}
\end{table}

\begin{table}
\begin{tabular}{|r|r|r|c|}\hline
Name & Length & Scop code & Family \\\hline \hline
1alc & 122 & 1004002001002 & 013 \\
1ctf & 68 & 1004026001001 & 001 \\
1cty & 108 & 1001003001001 & 004 \\
1fkb & 107 & 1004019001001 & 001 \\
1laa & 130 & 1004002001002 & 008 \\
1shg & 57 & 1002021002001 & 006 \\
1yeb & 108 & 1001003001001 & 004 \\
2fxb & 81 & 1004033001004 & 003 \\
351c & 82 & 1001003001001 & 017 \\
3il8 & 68 & 1004007001001 & 001 \\
\end{tabular}
\caption{Non redundant proteins used for test.}
\label{tab:test}
\end{table}

\begin{table}
\begin{tabular}{|r|r|r|}\hline
\multicolumn{3}{c||}{Fit cRMS (\AA)} \\ \hline $l=$4, m=10 & $l=$5,
 m=40 & $l=$6, m=100\\\hline \hline 1.06 $\pm$ 0.09 & 1.07 $\pm$ 0.12
 & 1.13 $\pm$ 0.11 \\
\end{tabular}
\caption{Results for the rigid fit procedure of the test proteins.}
\label{tab:rigfit}
\end{table}

\begin{table}
\begin{tabular}{|c|c|}\hline
Oligon rank& Forbidden position\\\hline \hline
7   & 3 \\
8   & 4 \\
10  & 3 \\
11  & 4 \\
13  & 3 \\
14  & 3 \\
16  & 3 \\
22  & 3 \\
24  & 4 \\
25  & 3 \\
27  & 3 \\ 
28  & 4 \\
29  & 4 \\
31  & 3 \\
32  & 4 \\ 
35  & 3 \\
35  & 4 \\
35  & 5 \\
38  & 3 \\\hline
\end{tabular}
\caption{List of the most significant forbidden occupations for
Proline on definite sites of oligons of length 5.}
\label{tab:forbidden}
\end{table}

\end{document}